\documentclass[%
 aip,
 jmp,%
 amsmath,amssymb,
 reprint,%
]{revtex4-1}

\usepackage{graphicx}
\usepackage{dcolumn}
\usepackage{bm}

\begin{document}

\preprint{AIP/123-QED}

\title[Moderate bandgap and high carrier mobility simultaneously realized in bilayer silicene by oxidation]{Moderate bandgap and high carrier mobility simultaneously realized in bilayer silicene by oxidation}

\author{Yan Qian}
 \email{qianyan@njust.edu.cn}
\author{Erjun Kan}
\author{Kaiming Deng}
\author{Haiping Wu}%
 \email{mrhpwu@njust.edu.cn}

\affiliation{
Department of Applied Physics, Nanjing University of Science and
Technology, Nanjing 210094, China
}%

\date{\today}

\begin{abstract}
Semiconductors simultaneously possessing high carrier mobility, moderate bandgap, and ambient environment stability are so important for the modern industry, and Si-based semiconducting materials can match well with the previous silicon based electronic components. Thus, searching for such Si-based semiconductors has been one hot project due to the lack of them nowadays. Here, with the help of density functional theory, we found that the oxidized bilayer silicene exhibits high carrier mobility with a moderate direct bandgap of 1.02 eV. The high carrier mobility is caused by the remaining of big $\pi$ bond, and the moderate bandgap is opened by the saturation of dangling Si 3\textit{p} bonds. Originated from the formation of strong Si-O and Si-Si bonds, the sample exhibits strong thermodynamic and dynamical stabilities. Our work indicates that the oxidized bilayer silicene has many potential applications in modern electronic fields.
\end{abstract}

\maketitle
\section{Introduction}
With the development of modern industry, devices with low power consumption and small scale are desirable in a wide range of applications, for instance,  optoelectronics, digital electronics, radio-frequency, chemical sensing, etc., and this encourages many researchers to search for corresponding materials. Low-dimensional materials, such as two-dimensional (2D) materials and so on, have been proved to be the potential ones that can satisfy the above demand\cite{Boon,Tang,Rao}. Moreover, in various electronic devices (field-effect transistors etc.), materials with semiconducting property are needed and essential. Therefore, low dimensional semiconductors show very important in such fields\cite{Li}, especially those with high carrier mobility, moderate bandgap, and ambient environment stability. However, such low dimensional semiconductors are limited as listed in the following. 2D atomically thin transition metal dichalcogenides (TMDs) MoS$_{2}$ has an enough wide bandgap of $\sim$1.8 eV\cite{Kin}, but it exhibits relatively low carrier mobility of $\sim$3.00 cm$^{2}$V$^{-1}$s$^{-1}$\cite{JD}. Black phosphorus has both enough wide bandgap of $\sim$ 2.0 eV and relatively high carrier mobility of 10$^{3}$ cm$^{2}$V$^{-1}$s$^{-1}$\cite{KOU,FEI,JIANG,LIU}, but it is unstable at ambient environment. Graphene possesses ultrahigh carrier mobility, but it behaves as a semimetal\cite{NKS,GM}, which motivates many researchers to tune its band structure. In 2018, an air-stable layered oxide semiconductor Bi$_{2}$O$_{2}$Se was experimentally reported as a promising one, it has a moderate bandgap of 0.8 eV and ultrahigh charge carrier mobility of 2.8$\times$10$^{5}$ cm$^{5}$V$^{-1}$s$^{-1}$\cite{Chen}.

However, most of the previous electronic components are made of silicon-based materials, and 2D silicon materials are the best ones that can match well with them. The above fact motivates researchers to focus their work on exploring 2D semiconducting silicon materials with high charge carrier mobility and moderate band gap. Silicene, a graphene-like single layer honeycomb structure of silicon, was the only experimentally synthesized one among the reported silicon elemental two-dimensional allotropes, and it was generated on Ag(111)\cite{BL,CHENL,FENG,PV}, ZrB$_{2}$(0001)\cite{AF}, Ir(111)\cite{GAO1}, Ru(0001)\cite{GAO2}, Pb(111)\cite{AS}, or ZrC(111)\cite{TA} substrates. The theoretical calculation reveals that the free-standing silicene exhibits linear electronic dispersion around the Dirac point, indicating that the electron can move massless. But it behaves as a semimetal with a zero energy gap, making researchers regulate its band structure in order to open an effective bandgap by diverse methods, such as surface adsorption\cite{QR,PZ,YD,DY,TPK,XW}, applying electric field\cite{NZY,DND}, multilayers\cite{WXQ,LJ,CQ}, substrate-inducing\cite{LCL,KY}, other 2D silicon allotropes\cite{SC0, HW}, and so on.

Furthermore, the free-standing silicene is unstable in ambient air environment due to its \textit{sp}$^{3}$ hybridization and dangling bonds, thus it is highly chemically active and easily to be oxidized in air environment. On the contrary, oxidization also have been used for tuning the physical properties of materials. As demonstrated by Du \textit{et al.}, silicene could be regulated from semimetal to semiconductor by chemically adsorbing oxygen atoms, and the bandgap could be tuned from 0.11 to 0.30 eV under different adsorption configurations and amounts of oxygen adatoms\cite{DY}. Gao \textit{et al.} also theoretically reported that three 2D oxidized silicon materials possess insulating property with bandgaps from 7.31 to 7.69 eV\cite{ZG}.

Although many semiconducting or insulating 2D silicon materials have been reported mainly via theoretical calculations, almost none of them possesses both moderate bandgap and high charge carrier mobility which are needed for the applications in FET. Here, we predicted that an oxidized bilayer silicene can possess both moderate bandgap and high carrier mobility. Du \textit{et al.} reported that silicene epitaxially grown on Ag(111) surface can fully oxidized\cite{DY}, but the structure exhibits a typical amorphous feature. This performance can be explained that Si-O-Si bonds are energetically stable due to the characteristic \textit{sp}$^{3}$ hybridization of Si, and the intrinsic structure of silicene will be destroyed when it is fully oxidized. Furthermore, the change of structure would heavily modify the properties of pure silicene. In 2016, Ritsuko \textit{et al.} successfully synthesized bilayer silicene after treating CaSi$_{2}$ with a BF$_{4}$$^{-}$-based ionic liquid\cite{RY}. Thus, if inserting oxygen atoms into bilayer silicene, both silicene layers can be oxidized with formation of Si-O-Si bonds as plotted in Fig. 1(a), and the construction of Si-O-Si bonds in the same silicene sublattice could be effectively avoided. As a result, the honeycomb structure in two silicene sublattices can be well retained. Following this idea, the oxidized bilayer silicene is explored by a first-principles theory in this work, and some interesting properties are unveiled.

\section{Computational methods}
The underlying ab initio structural relaxations and electronic band structure calculations are carried out in the framework of density functional theory (DFT) within generalized-gradient approximations using the PerdewBurke-Ernzerhof (PBE)\cite{Perdew} exchange correlation functional and projector-augmented-wave (PAW)\cite{Kresse1} potentials, as performed in VASP\cite{Kresse2}. To ensure high accuracy, the cutoff energy for plane waves of kinetic energy is up to 500 eV, and the vacuum space is 15 {\AA}. During geometry optimization, a 2$\times$2 supercell is employed and a 7$\times$7$\times$1 Monkhorstack grid is used in the first BZ, and all the atoms are allowed to be relaxed until the Hellmann-Feynman force on each atom converges to 0.01 eV/{\AA}.

\section{Results and discussions}
\begin{figure}[htbp]
\centering
\includegraphics[width=5.5cm]{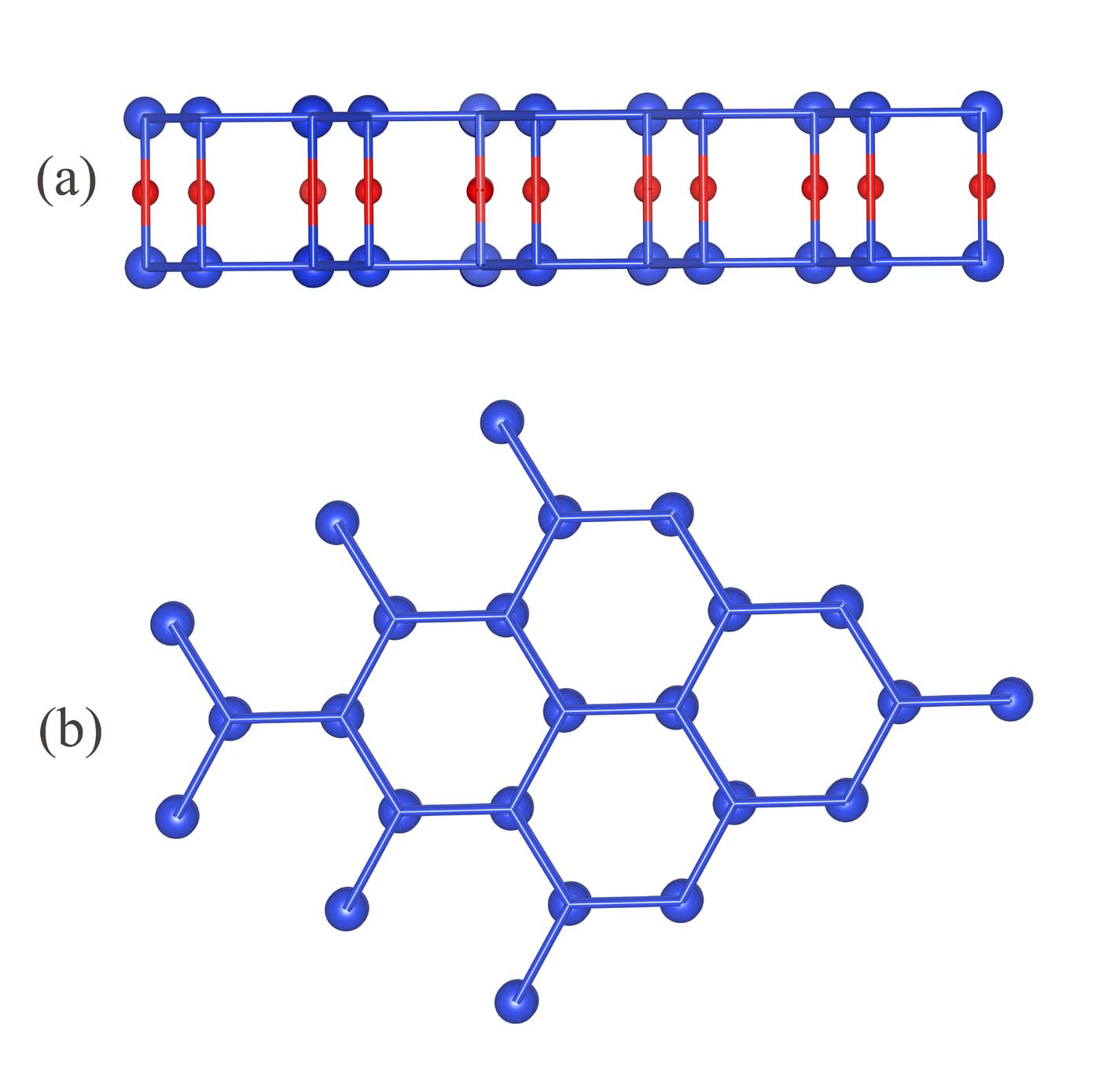}
\caption{(Color online). The optimized structure of oxidized bilayer silicene. (a) and (b) are the side and top views, respectively.}
\label{fig:Figure1}
\end{figure}
After inserted oxygen atoms between the two sublattice silicene layers, the honeycomb structure, instead of amorphous structure in oxidized monolayer silicene, is completely maintained in the bilayer silicene, as plotted in Fig. 1. Moreover, contrary to the buckling property in free-standing monolayer silicene, the oxidized bilayer silicene exhibits absolute plane characteristic. This structural difference is driven by the formation of strong Si-O bonds between O and two sublattice silicene layers, and the strong Si-O bonds can be confirmed by the electron localization function (ELF) in Fig. 2(a). The figure clearly shows there exists strong $\sigma$ bonds between the O and Si atoms, which suppresses the buckling deformation. This performance is the same as in the fully nitrogenated silicene\cite{YQ}, indicating the transition from \textit{sp}$^{3}$ to \textit{sp}$^{2}$ hybridization of Si orbitals. The ELF in Fig. 2(b) illuminates that the two nearest-neighboring Si atoms are bonded strongly as well, since there exhibit major $\sigma$ states. The strong bonds formed between the neighboring atoms indicates that the sample could exist stably.

The detail of the optimized structure is discussed in the following. The length of Si-O bonds is $\sim$ 1.65 {\AA}. This value is a bit longer than 1.61 or 1.62 {\AA} of Si-O bonds in the crystalline SiO$_{2}$ from experiment\cite{JP} or our calculational work, and is similar to the ones ranged from 1.63 to 1.66 {\AA} in the theoretically predicted 2D silicon dioxides\cite{ZG}. However, it is much shorter than 1.71 {\AA} in the oxidized monolayer silicene\cite{Liu}, implying the much stronger Si-O bonds in our sample. The Si-Si bonds are $\sim$ 2.49{\AA}, this value is larger than the theoretically predicted values of 2.25 or 2.28 {\AA} in the freestanding monolayer silicene\cite{SC,YQ} and 2.25 {\AA} in our previously predicted silicoctene\cite{HW}. This large change of Si-Si bond length can be explained by the following mechanism. With the intercalation of oxygen atoms in the bilayer silicene, Si atoms must provide electrons to form Si-O bonds, this results into the less electrons to form Si-Si bonds comparing with those in free-standing monolayer silicon allotropes. Besides, the electronegativity of O atom is much stronger than that of Si atom, this leads to the fact that there will be more electrons provided by Si atom to form O-Si bonds comparing with the formation of Si-Si bonds. As a result, the electrons forming Si-Si bonds in oxidized bilayer silicene are much smaller than those in the free-standing monolayer silicon materials. Additionally, Si atom prefers to \textit{sp}$^{3}$ hybridization, thus the hybridization transition of Si from \textit{sp}$^{3}$ to \textit{sp}$^{2}$ in this sample would cause the energetic decreasing of Si-Si bonds, which increases the Si-Si bond length.
\begin{figure}[htbp]
\centering
\includegraphics[width=8.5cm]{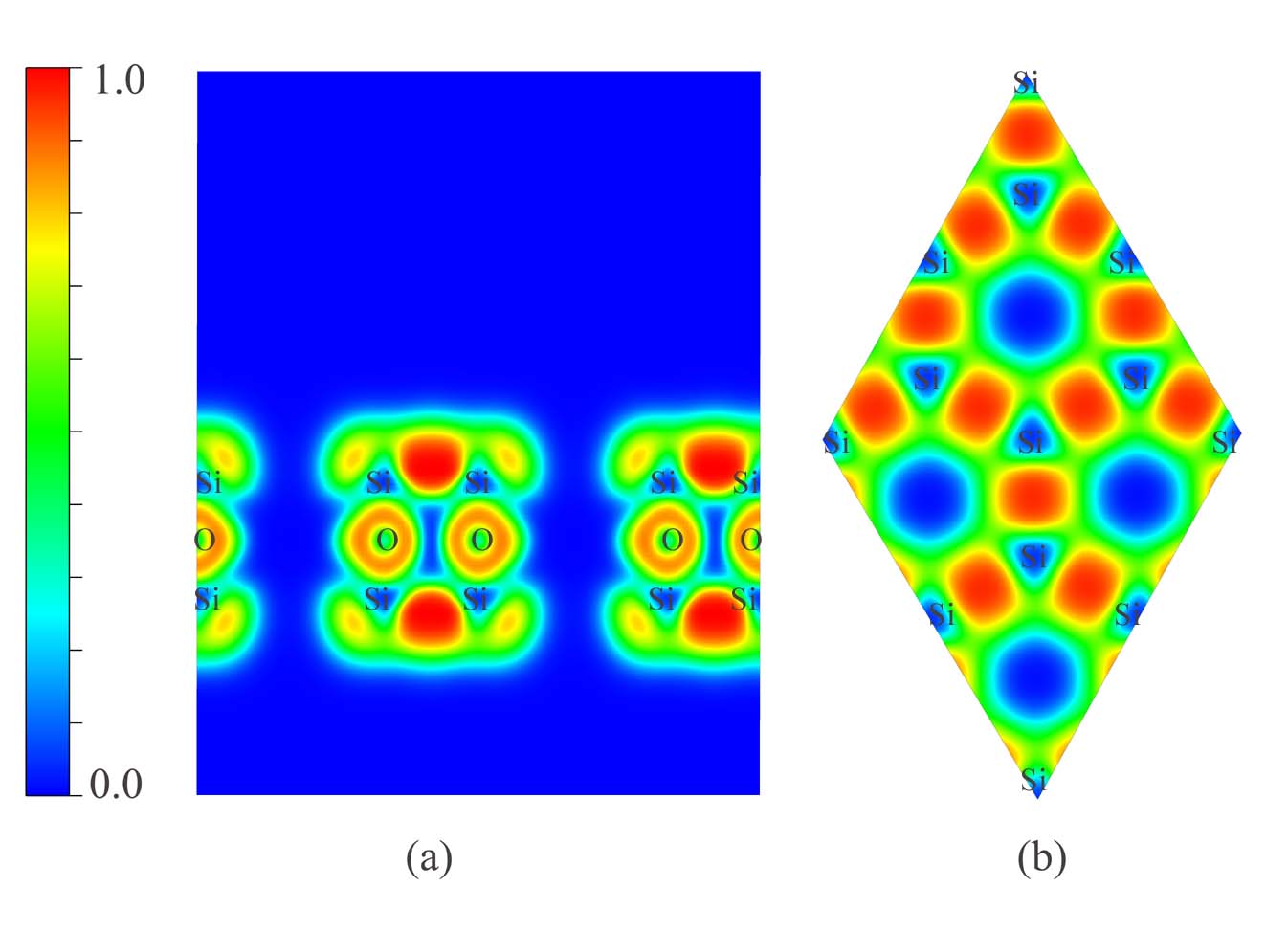}
\caption{(Color online). The calculated electron localization functions (ELF) of oxidized bilayer silicene. (a) and (b) are for (001) and (100) planes, respectively.}
\label{fig:Figure2}
\end{figure}

Furthermore, to explore the thermodynamic stability of the oxidized bilayer silicene, the adsorption energy (\textit{E}\textit{$_{ad}$}) is calculated, and it is expressed as:

\textit{E}$_{ad}$=(\textit{E}$_{total}$-\textit{2}\textit{E}$_{Silicene}$-\textit{n}\textit{E}$_{O}$)/\textit{n}$_{O}$

Here, \textit{{E}$_{total}$} and \textit{{E}$_{Silicene}$} are the total energies of the oxidized bilayer silicene and monolayer silicene, \textit{E}$_{O}$ is the energy per oxygen atom calculated from O$_{2}$ molecular in vacuum, and \textit{n} is the number of O atoms in the supercell. As a result, \textit{E}$_{ad}$ is $\sim$ 3.89 eV, this value is much larger than $\sim$ 2.40 eV in oxidized monolayer silicene\cite{Liu}. The larger \textit{E}$_{ad}$ demonstrates that the honeycomb structure of the oxidized bilayer silicene is much more stable than the structure of the oxidized monolayer silicene. Besides, this result is also consistent with the fact that Si-O bond length in the oxidized bilayer silicene is much shorter than that in the oxidized monolayer silicene.

The dynamical property is another important factor for the stability of materials, thus the phonon dispersion of oxidized bilayer silicene is studied as well, and the phonon dispersion curves are plotted in Fig. 3.
\begin{figure}[htbp]
\centering
\includegraphics[width=8.5cm]{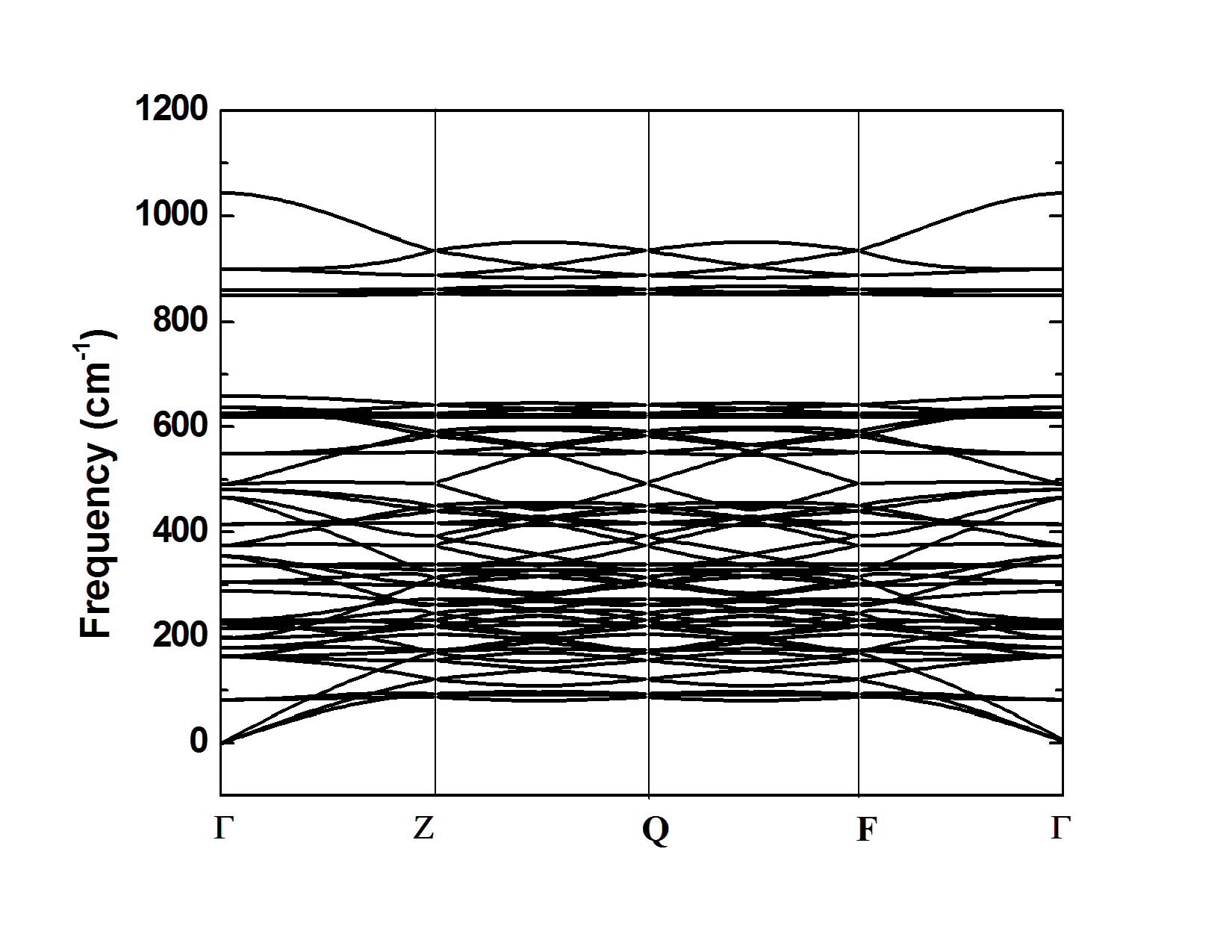}
\caption{The phonon dispersion curves for the oxidized bilayer silicene.}
\label{fig:Figure3}
\end{figure}
The figure clearly shows that there is no imaginary frequency for the structure, telling the sample is dynamically stable. The highest frequencies of the sample is $\sim$ 1050 cm$^{-1}$, similar with the value of $\gamma$-2D silica, a bit lower than the values of $\alpha$-2D silica and $\beta$-2D silica, and higher than $\delta$-2D silica\cite{ZG}. This performance can be explained by the fact that our sample and $\gamma$-2D silica almost have the same Si-O bond length of 1.65 {\AA}. Moreover, it is well known that higher frequency suggests the stronger bond and shorter bond length, and the trends of phonon frequency and Si-O bond lengthes in these 2D silicon oxides accords well with this rule, and it also indicates that theSi-O bonds in our sample are relatively strong.

The stability of the sample encourages us to investigate its electronic structure deeply. The density of electronic states (DOS) and band structure near the Fermi energy level (\textit{E}$_{F}$) are pictured in Figs. 4(a) and 4(b), respectively. The total DOS in Fig. 4(a) reveals that the sample behaves as a semiconductor, since there are no electronic states located at \begin{figure}[htbp]
\centering
\includegraphics[width=8.5cm]{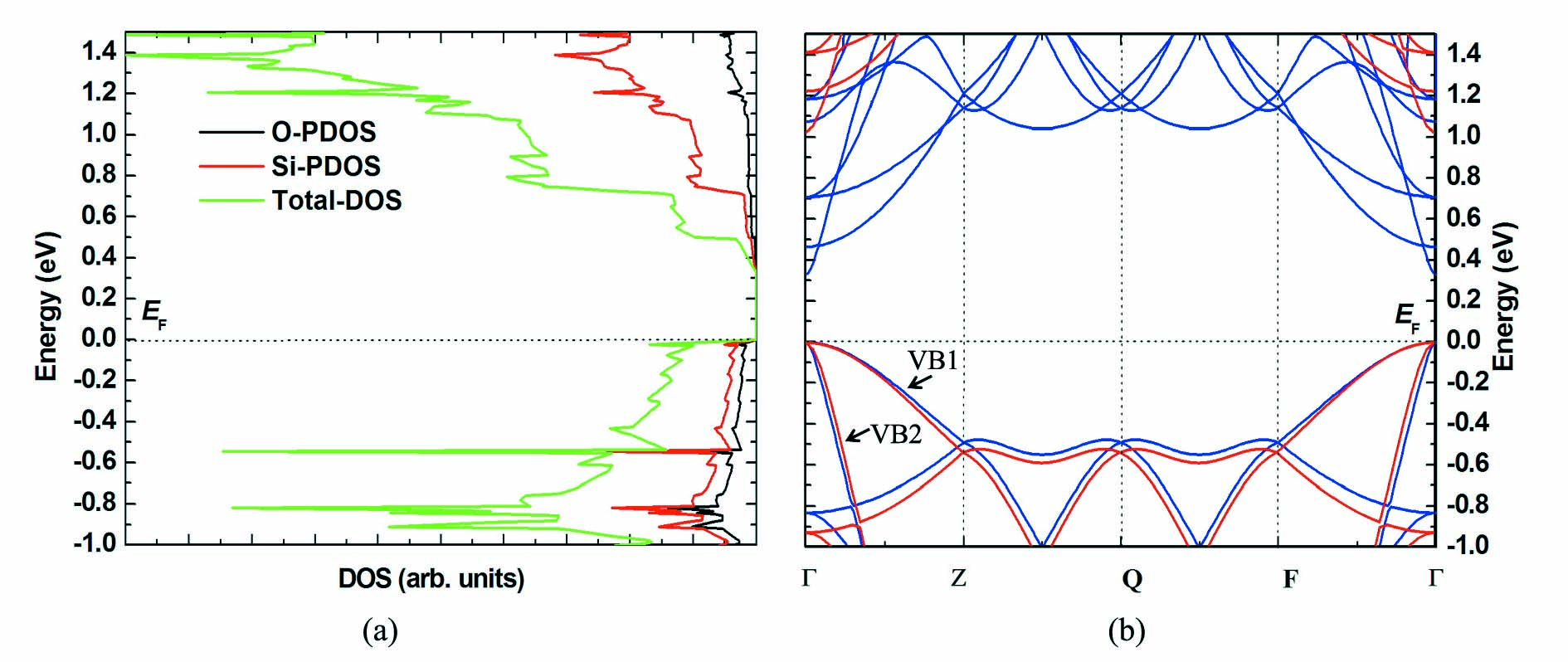}
\caption{(Color online). (a) The density of electronic states for the oxidized bilayer silicene. (b) The band structure for the oxidized bilayer silicene, and the blue and red lines are calculated by PBE and HSE functionals, respectively.}
\label{fig:Figure4}
\end{figure}
\textit{E}$_{F}$, and the energy gap is of $\sim$ 0.32 eV. The band structure as plotted in Fig. 4(b) shows that the conduction band minimum (CBM) and the valence band maximum (VBM) are both situated at the $\Gamma$ reciprocal point, demonstrating the character of a direct band gap semiconductor for the sample. At the $\Gamma$ reciprocal point, there exists a twofold degeneracy of the valence band. Since the PBE functional would usually underestimate the band gap, the screened hybrid functional of Heyd, Scuseria, and Ernzerhof (HSE), which can usually give more accurate band gaps than those from the PBE functional, is employed additionally, and the calculated band structure is plotted in Fig. 4(b) with red line. The direct band gap is $\sim$ 1.02 eV, this value is larger than $\sim$0.18 eV in oxidized momolayer silicene\cite{DY} and alkali atom absorbed silicene$\sim$0.50 eV\cite{QR}, and it is also larger than the indirect band gap of $\sim$ 0.55 eV in bilayer silicene\cite{WXQ}. The moderate band gap of the sample makes it suitable for the applications in FET device and so on.

For more detail, the partial density of electronic states (PDOS) near \textit{E}$_{F}$ is plotted in Fig. 4(a) as well. The result shows that there are many DOS peaks of Si and O atoms located at the same energy levels, indicating the strong coupling between the orbitals of oxygen and silicon atoms. This result is consistent with the ELF character that there exist strong Si-O bonds in the sample. PDOS also tells that both the valence and conduction bands are mainly originated from Si 3\textit{p} states, and minor are composed of O 2\textit{p} states. This result also can be proved by the partial (band decomposed) charge density of valence and conduction bands as plotted in Fig. 5. The figure
\begin{figure}[htbp]
\centering
\includegraphics[width=8.5cm]{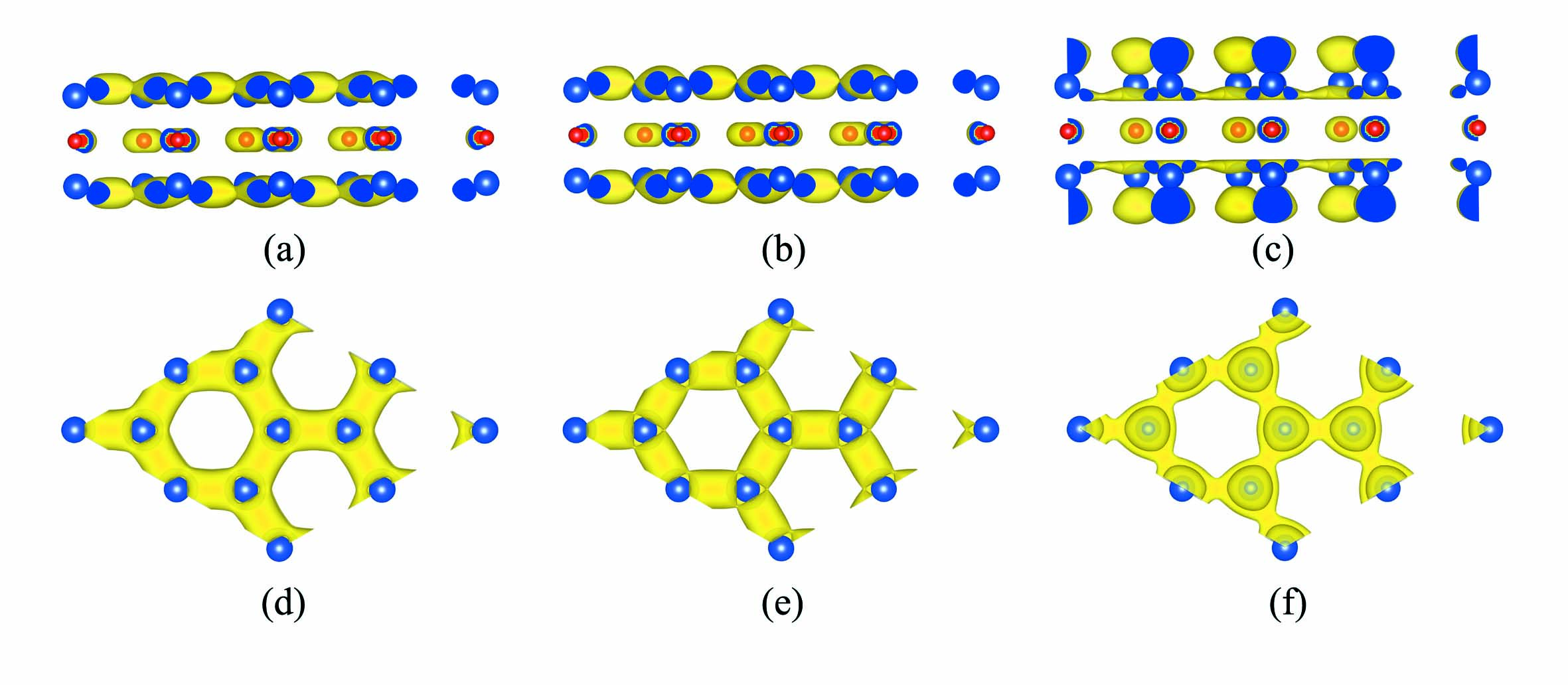}
\caption{(Color online). The band decomposed charge density of the two degenerated valence bands, (a) and (d) are for the VB2 valence band with heavy dispersion, and (b) and (e) for the VB1 valence band. The band decomposed charge density of the conduction band is in (c) and (f). (a), (b), and (c) are from the side view, and (d), (e), and (f) are from the top view.}
\label{fig:Figure5}
\end{figure}
shows that, for all the three bands including the two degenerated valence bands and one conduction band, the main charges are located at Si sites and minor ones are situated at O sites. Besides, big $\pi$ bonds are formed along silicon hexatomic ring for all these bands. The big $\pi$ bonds imply that the sample maybe has high carrier mobility. For the two degenerated valence bands, the figure shows that the big $\pi$ bonds in the VB2 band is stronger than that in the VB1 one, indicating the higher carrier mobility or lower carrier effective mass. This performance is consistent with the band dispersion feature in Fig. 4(b) that the dispersion of VB2 band is far stronger than that of VB1 band.

The charge carrier transport property is so important for the practical applications of semiconductors, thus the phonon limited carrier mobility of our sample is explored quantitatively finally. It is evaluated by the expression\cite{Bruzzone,Takagi,Fiori,Qiao,Jia}
\begin{equation}
\label{eq.3}
\mu_{2D}=\frac{e\hbar^{3}C_{2D}}{k_{B}Tm^{*}m_{a}(E_{l}^{i})^{2}}
\end{equation}
where \textit{e} is the electron charge and $\hbar$ is Planck's constant divided by 2$\pi$, \textit{k}$_{B}$ is Boltzmann's constant, and \textit{T} is the temperature with 300 \textit{K} employed here. m$^{*}$ is the effective mass of electron or hole at CBM or VBM in the transport direction, and it can be defined as 1/m$^{*}$=(1/$\hbar$$^{2}$)$\cdot$($\partial$$^{2}$E/$\partial$k$^{2}$). m$_{a}$ is the average effective mass determined by m$_{a}$ = (m$^{*}$$_{\Gamma-Z}$m$^{*}$$_{\Gamma-F}$)$^{1/2}$. The result shows that m$^{*}$ almost has the same values along the two directions. \textit{C}$_{2D}$ is the elastic modulus of the longitudinal strain in the propagation directions of the longitudinal acoustic wave and expressed by (\textit{E}-\textit{E}$_{0}$)/\textit{S}$_{0}$=\textit{C}$_{2D}$($\Delta$\textit{l}/\textit{l}$_{0}$)$^{2}$/2, here \textit{E} is the total energy and \textit{S}$_{0}$ is the lattice volume at equilibrium for a 2D system. \textit{E}$_{l}$$^{i}$ with the expression \textit{E}$_{l}$$^{i}$=$\Delta$\textit{V}$_{i}$/($\Delta$\textit{l}/\textit{l}$_{0}$) represents the deformation potential constant of VBM for hole or CBM for electron along the transport direction. Here $\Delta$\textit{V}$_{i}$ is the energy change of the \textit{i}$^{th}$ band under proper cell compression and dilatation, \textit{l}$_{0}$ and $\Delta$\textit{l} are the lattice constant and the deformation in the transport direction, respectively. The results show that the hole mobilities are $\sim$1.52$\times$10$^{4}$ cm$^{2}$V$^{-1}$s$^{-1}$ and $\sim$5.43$\times$10$^{5}$ cm$^{2}$V$^{-1}$s$^{-1}$, respectively, due to the twofold degeneracy of valence bands at the $\Gamma$ reciprocal point, and the electron mobility is $\sim$3.19$\times$10$^{4}$ cm$^{2}$V$^{-1}$s$^{-1}$. Notably, these values are much higher than those of Boron Nitride nanoribbons ($\sim$58.80 cm$^{2}$V$^{-1}$s$^{-1}$)\cite{HZ} and MoS$_{2}$ ($\sim$3.00 cm$^{2}$V$^{-1}$s$^{-1}$)\cite{JD}, and also higher than $\sim$10$^{3}$ cm$^{2}$V$^{-1}$s$^{-1}$ in atomically thin InSe\cite{Bandurin} and  $\sim$250.00 or $\sim$700.00 cm$^{2}$V$^{-1}$s$^{-1}$) for holes or electrons in Black Phosphorus at room temperature\cite{AN}. Such high carrier mobility illustrates that the oxidized bilayer silicene has great potentiality to be used in many applications, such as field-effect transistors, high efficiency solar cell, and so on.

\section{Conclusions}
In summary, a bilayer silicene could be tuned into a semiconductor with both moderate bandgap and high carrier mobility by oxidization. With the insertion of O atoms between the two silicene sublayers, the dangling Si 3\textit{p} bonds are saturated by the formation of strong Si-O bonds, while the big $\pi$ bonds around the Si hexatomic ring are remained. The above special properties originate from the maintaining of the honeycomb structure.

\vspace{1ex}
\acknowledgments
This work was supported by the National Natural Science Foundation of China (Grant no. 11404168).


\begin{thebibliography}{99}
\bibitem{Boon} K. T. Boon and Sun X. H., Chem. Rev. {\bf 107}, 1454 (2007).
\bibitem{Tang} Q. Tang and Z. Zhou, Prog. Mater. Sci. {\bf 58}, 1244 (2013).
\bibitem{Rao} C. N. R. Rao, K. Gopalakrishnan, and U. Maitra, ACS Appl. Mater. Interfaces. {\bf 7}, 7809 (2015).
\bibitem{Li} M. Y. Li, S. K. Su, H. S. Philip Wong, and L. J. Li, Nature {\bf 567}, 169 (2019).
\bibitem{Kin} K. F. Mak, C. G. Lee, J. Hone, J. Shan, and T. F. Heinz, Phys. Rev. Lett. {\bf 105}, 136805 (2010).
\bibitem{JD} K. S. Novoselov, D. Jiang, F. Schedin, T. J. Booth, V. V. Khotkevich, S. V. Morozov, and A. K. Geim, Proc. Natl. Acad. Sci. U.S.A. {\bf 102}, 10451 (2005).
\bibitem{KOU} L. Z. Kou, T. Frauenheim, and C. F. Chen, J. Phys. Chem. Lett. {\bf 5}, 2675 (2014).
\bibitem{FEI} R. Fei and L. Yang, Nano Lett. {\bf 14}, 2884 (2014).
\bibitem{JIANG} J. W. Jiang and H. S. Park, Nat. Commun. {\bf 5}, 4727 (2014).
\bibitem{LIU} Q. Liu, X.Zhang, L. B. Bdalla, A. Fazzio, and A. Zunger, Nano Lett. {\bf 15}, 1222 (2015).
\bibitem{NKS} K. S. Novoselov, A. K. Geim, S. V. Morozov, D. Jiang, M. I. Katsnelson, I. V. Grigorieva, S. V. Dubonos, and A. A. Firsov,  Nature {\bf 438}, 197 (2005).
\bibitem{GM} A. K. Geim and K. S. Novoselov, Nat. Mater. {\bf 6}, 183 (2007).
\bibitem{Chen} C. Chen, M. X. Wang, J. X. Wu, H. X. Fu, H. F. Yang, Z. Tian, T. Tu, H. Peng, Y. Sun, X. Xu, J. Jiang, N. B. M. Schr\"{o}ter, Y. W. Li, D. Pei, S. Liu, S. A. Ekahana, H. T. Yuan, J. M. Xue, G. Li, J. F. Jia, Z. K. Liu, B. H. Yan, H. L. Peng, Y. L. Chen, Sci. Adv. {\bf 4}, eaat8355 (2018).
\bibitem{BL} B. Lalmi, H. Oughaddou, H. Enriquez, A. Kara, S. Vizzini, B. Ealet, and B. Aufray, Appl. Phys. Lett. {\bf 9}, 223109 (2010).
\bibitem{CHENL} L. Chen, C. C. Liu, B. J. Feng, X. Y. He, P. Cheng, Z. J. Ding, S. Meng, Y. G. Yao, and K. H. Wu, Phys. Rev. Lett. {\bf 109}, 056804 (2012).
\bibitem{FENG} B. Feng, Z. Ding, S. Meng, Y. Yao, X. He, P. Cheng, L. Chen, and K. Wu, Nano Lett. {\bf 12}, 3507 (2012).
\bibitem{PV} P. Vogt, P. D. Padova, C. Quaresima, J. Avila, E. Frantzeskakis, M. C. Asensio, A. Resta, B. Ealet, and G. L. Lay, Phys. Rev. Lett. {\bf 108}, 155501 (2012).
\bibitem{AF} A. Fleurence, R. Friedlein, T. Ozaki, H. Kawai, Y. Wang, and Y. Yamada-Takamura, Phys. Rev. Lett. {\bf 108}, 245501 (2012).
\bibitem{GAO1} L. Meng, Y. L. Wang, L. Z. Zhang, S. X. Du, R. T. Wu, L. F. Li, Y. Zhang, G. Li, H. T. Zhou, W. A. Hofer, H. J. Gao, Nano Lett. {\bf 13}, 685 (2013).
\bibitem{GAO2} L. Huang, Y. F. Zhang, Y. Y. Zhang, W. Y. Xu, Y. D. Que, E. Li, J. B. Pan, Y. L. Wang, Y. Q. Liu, S. X. Du, S. T. Pantelides, H. J. Gao, Nano Lett. {\bf 17}, 1161 (2017).
\bibitem{AS} A. Stepniak-Dybala and M. Krawiec, J. Phys. Chem. C {\bf 123}, 17019 (2019).
\bibitem{TA} T. Aizawa, S. Suehara, and S. Otani, J. Phys. Chem. C {\bf 118}, 23049 (2014).
\bibitem{QR} R. Quhe, R. X. Fei, Q. H. Liu, J. X. Zheng, H. Li, C. Y. Xu, Z. Y. Ni, Y. Y. Wang, D. P. Yu, Z. X. Gao, and J. Lu, Sci. Rep. {\bf 2}, 853 (2012).
\bibitem{PZ} P. Zhang, X. D. Li, C. H. Hu, S. Q. Wu, and Z. Z. Zhu, Phys. Lett. A {\bf 376}, 1230 (2012).
\bibitem{YD} Y. Ding and Y. Wang, Appl. Phys. Lett. {\bf 100}, 083102 (2012).
\bibitem{DY} Y. Du, I. C. Zhuang, H. S. Liu, X. Xu, S. Eilers, K. H. Wu, P. Cheng, J. J. Zhao, X. D. Pi, K. W. See, G. Peleckis, X. L. Wang, and S. X. Dou, ACS Nano {\bf 8}, 10019 (2014).
\bibitem{TPK} P. K. Thaneshwor, S. Georg, and S. F. Michael, J. Phys. Chem. C {\bf 118}, 23361 (2014).
\bibitem{XW} X. Wang, H. Liu, and S. T. Tu, RSC Adv. {\bf 5}, 6238 (2015).
\bibitem{NZY} Z. Y. Ni, Q. H. Liu, K. C. Tang, J. X. Zheng, J. Zhou, R. Qin, Z. X. Gao, D. P. Yu, and J. Lu, Nano Lett. {\bf 12}, 113 (2012).
\bibitem{DND} N. D. Drummond, V. Z\'{o}lyomi, and V. I. Falko, Phys. Rev. B {\bf 85}, 075423 (2012).
\bibitem{WXQ} X. Q. Wang and Z. G. Wu, Phys. Chem. Chem. Phys. {\bf 19}, 2148 (2017).
\bibitem{LJ} J. Lv, M. L. Xu, S. R. Line, X. C. Shao, X. Y. Zhang, Y. H. Liu, Y. C. Wang, Z. F. Chen, and Y. M. Ma, Nano Energy {\bf 51}, 489 (2018).
\bibitem{CQ} C. Qian and Z. Li, Comp. Mater. Sci. {\bf 172}, 109354 (2020).
\bibitem{LCL} C. L. Lin, R. Arafune, K. Kawahara, M. Kanno, N. Tsukahara, E. Minamitani, Y. Kim, M. Kawai, and N. Takagi, Phys. Rev. Lett. {\bf 110}, 076801 (2013).
\bibitem{KY} K. Yang, W. Q. Huang,  W. Y. Hu, G. F. Huang, and S. C. Wen, Nanoscale, {\bf 10}, 14667 (2018).
\bibitem{SC0} S. Cahangirov, V. Ongun\"{O}z\c{c}lik, A. Rubio, and S. Ciraci, Phys. Rev. B {\bf 90}, 085426 (2014).
\bibitem{HW} H. P. Wu, Y. Qian, Z. W. Du, R. Z. Zhu, E. J. Kan, and K. M. Deng, Phys. Lett. A {\bf 381}, 3754 (2017).
\bibitem{ZG} Z. B. Gao, X. Dong, N. B. Li, and J. Ren, Nano Lett. {\bf 17}, 772 (2017).
\bibitem{RY} R. Yaokawa, T. Ohsuna, T. Morishita, Y. Hayasaka, M. J. S. Spencer, and H. Nakano, Nat. Commun. {\bf 7}, 10657 (2016).
\bibitem{Perdew} J. P. Perdew, K. Burke, and M. Ernzerhof, Phys. Rev. Lett. {\bf 77}, 3865 (1996).
\bibitem{Kresse1} G. Kresse and D. Joubert, Phys. Rev. B: Condens. Matter {\bf 59}, 1758 (1999).
\bibitem{Kresse2} G. Kresse and J. Furthm\"{u}ller, Comput. Mater. Sci. {\bf 6}, 15 (1996).
\bibitem{YQ} Y. Qian, H. P. Wu, E. J. Kan, R. F. Lu, and K. M. Deng, J. Appl. Phys. {\bf 120}, 234303 (2016).
\bibitem{JP} J. J. Pluth, J. V. Smith, and J. Faber Jr,  J. Appl. Phys. {\bf 57}, 1045 (1985).
\bibitem{Liu} G. Liu, X. L. Lei, M. S. Wu, B. Xu, and C. Y. Ouyang, EPL {\bf 106}, 47001 (2014).
\bibitem{SC} S. Cahangirov, M. Topsakal, E. Akt\"{u}rk, H. \c{S}ahin, and S. Ciraci, Phys. Rev. Lett. {\bf 102}, 236804 (2009).
\bibitem{Bruzzone} Bruzzone S. and Fiori G. Appl. Phys. Lett. {\bf 99}, 222108 (2011).
\bibitem{Takagi}  S. Takagi, A. Toriumi, M. Iwase, and H. Tango, IEEE Trans. Electr. Dev. {\bf 41}, 2357 (1994).
\bibitem{Fiori} G. P. Fiori, IEEE {\bf 101}, 1653 (2013).
\bibitem{Qiao} J. Qiao, X. Kong, Z. X. Hu, F. Yang, and W. Ji, Nat. Commun. {\bf 5}, 4475 (2014).
\bibitem{Jia} J. F. Xie, Z. Y. Zhang, D. Z. Yang, D. S. Xue, and M. S. Si, J. Phys. Chem. Lett. {\bf 5}, 4073 (2014).
\bibitem{HZ} H. Zeng, C. Zhi, Z. Zhang, X. Wei, X. Wang, W. Guo, Y. Bando, and D. Golberg, Nano Lett. {\bf 10}, 5049 (2010).
\bibitem{Bandurin} D. A. Bandurin, A. V. Tyurnina, G. L. Yu, and A. Mishchenko, Nat. Nanotech. {\bf 12}, 223 (2017).
\bibitem{AN} A. N. Rudenko, S. Brener, and M. I. Katsnelson, Phys. Rev. Lett. {\bf 116}, 246401 (2016).




\end{thebibliography}
\end{document}